\documentclass[
  aps,  
  twocolumn,
  prl,
  superscriptaddress,
  floatfix
]{revtex4}

\usepackage{amssymb,amsfonts,amsmath,dsfont}
\usepackage{times} 
\usepackage{graphicx} 
\usepackage{gensymb}
\usepackage{pifont}
\usepackage{bm} 
\usepackage{color}
\usepackage{accents}
\usepackage{mathtools}
\usepackage{hyperref}
\usepackage{url}
\hypersetup{colorlinks=true, citecolor=blue, urlcolor=blue, linkcolor=blue}
\usepackage{multirow}
 
\makeatletter
\newcommand*{\supplementarystart}{%
  \close@column@grid%
  \clearpage%
  \onecolumngrid%
  \setcounter{enumiv}{0} % resets counter for references
  \setcounter{equation}{0} % resets counter for equations
  \setcounter{figure}{0} % resets counter for figs
  \setcounter{table}{0} % resets counter for tables
  \setcounter{page}{1}
  \c@secnumdepth=4
  \renewcommand{\theequation}{s\arabic{equation}} % equations numbered with S...
  \renewcommand{\bibnumfmt}[1]{[s##1]} % bibtems [S...]
  \renewcommand{\@onlinecite}{s\citealp} % citations [S...]
  \renewcommand{\cite}[1]{{[}\onlinecite{##1}{]}}
  \renewcommand{\thefigure}{s\arabic{figure}}
  \renewcommand{\thetable}{s\Roman{table}}
  \renewcommand{\thepage}{s\arabic{page}}
}
\makeatother
 
\newcommand{\s}{\sum\limits} 
 
\newcommand{\pa}{\partial} 
 
\newcommand{\be}{\begin{equation}} 
\newcommand{\e}{\end{equation}} 
\newcommand{\beml}{\begin{subequations}} 
\newcommand{\eml}{\end{subequations}} 
\newcommand{\beq}{\begin{eqnarray}} 
\newcommand{\eq}{\end{eqnarray}} 
\newcommand{\ba}{\begin{array}} 
\newcommand{\ea}{\end{array}} 
\newcommand{\bpm}{\begin{pmatrix}} 
\newcommand{\epm}{\end{pmatrix}} 
\newcommand{\bc}{\begin{cases}} 
\newcommand{\ec}{\end{cases}} 
\newcommand{\lt}{\left} 
\newcommand{\rt}{\right} 
\newcommand{\n}{\nonumber} 
\newcommand{\la}{\langle} 
\newcommand{\ra}{\rangle}

\newcommand{\bb}{\boldsymbol} 
\newcommand{\bbm}{\mathbf}

\begin{document}
	
\title{Signatures of quartic asymmetric exchange in a class of two dimensional magnets}	

\author{G.~Rakhmanova}
\affiliation{ITMO University, Faculty of Physics, Saint-Petersburg, Russia}
\author{A.~Osipov}
\affiliation{ITMO University, Faculty of Physics, Saint-Petersburg, Russia}
\author{D.~Ilyin}
\affiliation{ITMO University, Faculty of Physics, Saint-Petersburg, Russia}
\author{I.~Shushakova}
\affiliation{ITMO University, Faculty of Physics, Saint-Petersburg, Russia}
\author{I.\,A.~Ado}
\affiliation{Institute for Theoretical Physics, Utrecht University, 3584 CC Utrecht, The Netherlands}
\affiliation{Institute for Molecules and Materials, Radboud University, 6525 AJ Nijmegen, The Netherlands}	
\author{I.~Iorsh}
\affiliation{ITMO University, Faculty of Physics, Saint-Petersburg, Russia}	
\author{M.~Titov} 
\affiliation{Institute for Molecules and Materials, Radboud University, 6525 AJ Nijmegen, The Netherlands}	

\date{\today}

\begin{abstract}
Indirect quartic interaction of spins is suggested to play an important role in two dimensional magnets with trigonal prismatic symmetry such as Fe$_3$GeTe$_2$ monolayer. The proposed interaction is described by terms in micromagnetic energy that are linear in magnetization gradients. Such terms enable the stability of non-collinear magnetic textures. We investigate signatures of the quartic interaction in magnon spectra in the presence of anisotropy and external magnetic field. We also show how magnetic spirals, which are induced by the proposed interaction, can be manipulated by external field. Our analysis is based on symmetry considerations and can be used to quantify the quartic interaction strength in experiments with magnetic monolayers.
\end{abstract}
	
\maketitle

Spin orbit coupling in magnetic insulators is known to induce antisymmetric exchange of neighboring spins -- the so-called Dzyalosinskii-Moria interaction (DMI) \cite{Moriya1960,Dzyaloshinsky1958}. In conducting magnets as well as in heavy-metal/ferromagnet heterostructures and multilayers the role of DMI is played by a long-range indirect antisymmetric exchange \cite{FertLevy1980,Hiroshi2003,Pyatakov2014,AdoDMI2018}. In both cases, the spin-orbit induced interactions are represented in the micromagnetic energy of a magnet by terms that are linear in magnetization gradients. The latter is deemed responsible for the formation of non-collinear magnetic textures such as magnetic helixes, conical spirals and skyrmion crystals in a variety of magnetic materials \cite{Ishikawa1976,Shirane1983,Muhlbauer2009,Grigoriev2009,Moreau-Luchaire2016,Woo2016}. Thus, from theory point of view, the presence of linear-in-gradient terms play the leading role in stabilizing non-collinear magnetism.

From one hand, manipulating non-collinear magnetic textures by optical or electrical means becomes a mainstream for the development of ultra-fast spintronics devices \cite{Parkin-race-track-2008,yoshida2012,Menzel_PRL2012,Fert2013skyrmion-racetrack,Wiesendanger2013,Steinbrecher2018}. From the other hand, high-quality two dimensional magnets offer diverse playground to enable such a development \cite{Novoselov2019}. It is, therefore, important to investigate the origin of non-collinear magnetism in monocrystalline 2D magnets.

In this paper we focus on a class of 2D ferromagnetic materials, for example Fe$_3$GeTe$_2$ monolayers, that have an atomic lattice with trigonal prismatic symmetry of D$_\textrm{3h}$ point group. This is effectively the honeycomb magnetic lattice with non-equivalent A and B sub-lattices. Such a lattice is characterized by a broken inversion that should lead to strong DMI. 

It has been shown, however, that for lattices with D$_\textrm{3h}$, C$_\textrm{3h}$, and T$_\textrm{d}$ point group symmetries the DMI cannot induce linear-in-gradient terms in micromagnetic energy \cite{AdoDMI2020} and, moreover, such terms are generally forbidden in quadratic order with respect to magnetization. Consequently, in the corresponding materials the role of DMI (and any other long-range two-spin antiymmetric exchange) must be negligible for the formation of non-collinear magnetic textures that are smooth on atomic scales \cite{Hals2019,ManchonFGT2020,D3h2021}. 

Despite this conclusion, experiments in Fe$_3$GeTe$_2$ demonstrate non-collinear order with periods of hundreds of nanometers up to a monolayer limit \cite{WangFGTskyrmion2020,MarcosFGTspiral2020,ParkFGTskyrmion2021}. Such observations may indicate the importance of other magnetic interactions that involve more than two spins.   

Indeed, the lattices with D$_\textrm{3h}$, C$_\textrm{3h}$, and T$_\textrm{d}$ point groups allow for quartic spin interaction that leads to linear-in-gradient terms in micromagnetic energy. We coin the term quartic asymmetric exchange to refer to such interaction. The proposed interaction differs from various multi-spin generalizations of DMI that have been recently considered \cite{Laszloffy_Multispin2019,Brinker_2019,Hoffman2020}. 

The quartic asymmetric exchange is known to stabilize conical magnetic spirals \cite{AdoTd2021,Rybakov2021,D3h2021}. The aim of this paper is to investigate how this interaction can be detected and quantified in 2D ferromagnetic monolayer of D$_\textrm{3h}$ point group.

\begin{figure}[t]
\centerline{\includegraphics[width=0.8\columnwidth]{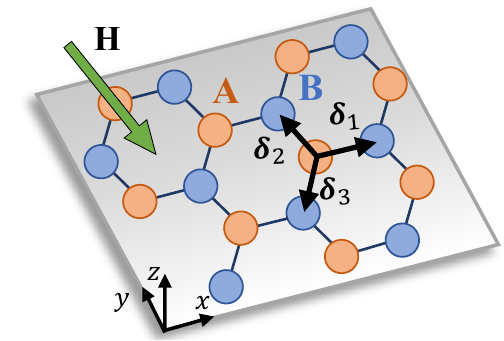}} 
\caption{Schematic illustration of the magnetic honeycomb lattice for a material with D$_\textrm{3h}$ point group. The vectors $\bb{\delta}_\alpha$ are defined as unit vectors from an atom on the sub-lattice $A$ in the direction of the three nearest neighbors on the sub-lattice B. The $x$ axis is chosen along the armchair direction.}
\label{fig:lattice}
\end{figure}

Quartic asymmetric exchange for 2D ferromagnets with trigonal prismatic symmetry has been first introduced in Ref.~\cite{D3h2021}. Independently of its microscopic origin it is represented by a unique term in the energy density of a magnet, 
\be 
\label{chiral}
w_{4S}\propto (\bbm{n}\cdot \bb{\delta}_1)(\bbm{n}\cdot \bb{\delta}_2)(\bbm{n}\cdot \bb{\delta}_3)\,\bb{\nabla}\cdot\bbm{n},
\e
where the unit vector field $\bbm{n}(\bb{r})$ describes the direction of local magnetization as the function of two-dimensional coordinate $\bbm{r}$, while the three vectors $\bb{\delta}_\alpha$ connect neighboring cites on the honeycomb magnetic lattice as shown in Fig.~\ref{fig:lattice}. 

%%%%%%%%%%%%%%%%%%%%%%%%%%%%
%%%% fig:lattice
%%%%%%%%%%%%%%%%%%%%%%%%%%%%
%%%%%%%%%%%%%%%%%%%%%%%%%%%%%

We shall assume that ferromagnetic order is only weakly perturbed by such symmetry-breaking interaction, hence the local magnetic moments vary smoothly on atomic scales and the continuous limit description is justified. 

The interaction of the type of Eq.~(\ref{chiral}) can be obtained from a four-spin interaction of the nearest neighbor spins for a Heisenberg model on a honeycomb lattice by making expansion with respect to magnetization gradients. Such lattice interaction has the form
\begin{align}
I_{4S}\propto &\; S_i^{\textrm{A}(1)} S_i^{\textrm{A}(2)} S_i^{\textrm{A}(3)} \lt(S_{j_1}^{\textrm{B}(1)}+S_{j_2}^{\textrm{B}(2)}+S_{j_3}^{\textrm{B}(3)}\rt) \n\\ 
&-  S_j^{\textrm{B}(1)} S_j^{\textrm{B}(2)} S_j^{\textrm{B}(3)} (S_{i_1}^{\textrm{A}(1)}+S_{i_2}^{\textrm{A}(2)}+S_{i_3}^{\textrm{A}(3)}),
\label{quartic0}
\end{align}
where $S^{(\alpha)}=\bbm{S}\cdot\bb{\delta}_\alpha$ denotes spin projection on one of the three possible nearest neighbor directions. The vector $\bbm{S}_{i}^\textrm{A}$ refers to a spin on the site $i$ of the sub-lattice A that is surrounded by three neighboring sites: $j_1$, $j_2$ and $j_3$ of the sub-lattice B. The neighboring sites $j_\alpha$ are obtained by the translations along the vectors $\bb{\delta}_\alpha$ as shown in Fig.~\ref{fig:lattice}. Thus, the product of spin projections on an A site interacts with the corresponding spin projections on the three neighboring B sites. The corresponding product of B spins interacting with A neighbors comes with an opposite sign. Such a sign alteration between A and B sub-lattices reflects a sub-lattice ``asymmetry'' of the quartic exchange. 

The interaction of the type of Eq.~(\ref{quartic0}) summed over the entire lattice can be compactly written as
\begin{align}
I_{4S}&\propto \s_{\la i, j\ra}  \lt[\overline{S}_i^{\textrm{A}} S_{j_\alpha}^{\textrm{B}(\alpha)} - \overline{S}_j^{\textrm{B}} S_{i_\alpha}^{\textrm{A}(\alpha)}\rt],
\label{quartic1}
\end{align}
where $\overline{S}_i ={S}_i^{(1)}{S}_i^{(2)}{S}_i^{(3)}$, and ${S}_i^{(\alpha)} =\bbm{S}_i\cdot\bb{\delta}_\alpha$. Note that even though the vectors $\bb{\delta}_\alpha$ are defined specifically as vectors from A to B sites, their signs can be reversed without any effect on the interaction. Hence the opposite definition as vectors from B to A is equally possible. 

Quartic interaction of the type of Eq.~(\ref{quartic1}) has nothing to do with various DMI generalizations: $\bb{D}_{ijlk}\,\cdot \bbm{S}_i\times\bbm{S}_j(\bbm{S}_l\cdot \bbm{S}_k)$. Instead, the asymmetry in Eq~(\ref{quartic1}) is not of the vector product type but arises from a non-equivalence of A and B sub-lattices. 

The contribution of Eq.~(\ref{chiral}) is expected to be especially strong in conducting ferromagnets where it is enabled by conduction electrons \cite{FertLevy1980,Hiroshi2003,Pyatakov2014,AdoDMI2018}. In this case, its lattice representation is essentially long range. Still, the short range interaction of Eq.~(\ref{quartic1}) may be of use for efficient computer analysis of the effects induced by quartic asymmetric exchange.

If we chose $x$ axis along the armchair direction of the honeycomb lattice as shown in Fig.~\ref{fig:lattice} and define $\bb{\delta}_1=(1,0)$, $\bb{\delta}_{2,3}=-(1/2, \pm\sqrt{3}/2)$, we obtain
\be
w_{4S}=n_x(n_x^2-3n_y^2)(\partial_x n_x +\partial_y n_y).
\e
Thus, the free energy functional $F[\bbm{n}]$ for a D$_\textrm{3h}$ ferromagnet is given by $F=\int d^2\bbm{r}\, f(\bbm{n})$ with the energy density
\be
f(\bbm{n})=A\,[\lt(\pa_x \bbm{n}\rt)^2+\lt(\pa_y \bbm{n}\rt)^2]+K\,n_z^2+8\mathcal{D}\,w_{4S}-\bbm{H}\cdot\bbm{n},
\e
where $A, K, \mathcal{D}$ are the stiffness, anisotropy and quartic interaction coefficients, respectively. 

Following Ref.~\cite{D3h2021} we choose to minimize the free energy for a generic conical spiral 
\be
\bbm{n}(\bbm{r})=\bbm{m}\cos\alpha +[\bbm{m}_{\theta}\cos \bbm{k}\cdot\bbm{r}+\bbm{m}_{\varphi}\sin \bbm{k}\cdot\bbm{r}]\sin\alpha,
\e
where $\bbm{m}$ is cone axis which is the unit vector in the direction of average magnetization, while $\bbm{m}_{\theta}$ and $\bbm{m}_{\varphi}$ are mutually perpendicular vectors that are both perpendicular to $\bb{m}$. Opening of the magnetization cone is governed by the angle $\alpha$. For $\alpha=0$ one finds the collinear state $\bb{n}(\bbm{r})=\bbm{m}$.

Below we define $\bbm{m}=(\sin\theta\cos\varphi,\sin\theta\sin\varphi,\cos\theta)$,  $\bbm{m}_{\varphi}=(-\sin\varphi,\cos\varphi,0)$ and $\bbm{m}_{\theta}=[\bbm{m}\times\bbm{m}_{\varphi}]$. Thus, the angle $\varphi$ specifies the direction of the in-plane component of the cone axis $\bb{m}$. 

The vector $\bbm{k}=(k_x, k_y)$ is the wave vector of the spiral. In the thermodynamic limit the free energy must not depend on a spacial translation along $\bbm{k}$, $\bbm{k}\cdot\bbm{r}\to \bbm{k}\cdot\bbm{r}+\gamma$. Averaging over such translations, which is equivalent to the averaging over the angle $\gamma$, results in a function $\la f \ra$ that depends only on the spiral parameters $\bbm{k}$, $\alpha$, $\varphi$ and $\theta$. Minimization of $\la f \ra$ with respect to these parameters is a tractable problem that can be approached analytically. 

Indeed, the minimization with respect to the wave vector is readily performed since $\la f \ra$ has a quadratic dependence on $\bbm{k}$. The minimum is achieved for
\be
\begin{pmatrix} k_x \\ k_y \end{pmatrix}= -\frac{3\mathcal{D}}{A}(5\cos^2\alpha-1)\sin^2\theta\cos\theta \begin{pmatrix}\sin 2\varphi \\ \cos 2\varphi
\end{pmatrix}.
\e
From this result one can immediately see that 
\be
\frac{\bbm{k}\cdot(\hat{z}\times\mathbf{m})}{k |\hat{z}\times\mathbf{m}|}=\cos 3\varphi,\quad \mbox{where}\; k=|\bbm{k}|.
\label{eq:dir}
\e
Thus, the direction of the spiral wave-vector $\bbm{k}$ is determined by the direction of the magnetic cone axis in-plane projection $\bbm{m}_{\|}=\sin\theta(\cos\varphi,\sin\varphi,0)$ with respect to the lattice. 

For $\bbm{m}_{\|}$ directed along $x$ axis ($\varphi=0$), the spiral propagates along $y$ axis, hence $\bbm{k}$ and $\bbm{m}_{\|}$ are perpendicular vectors. For  $\bbm{m}_{\|}$ directed along the zigzag direction of the honeycomb lattice ($\varphi=\pi/6$), the vectors $\bbm{k}$ and $\bbm{m}_\parallel$ are parallel to each other \cite{D3h2021}. 

Consequently, one can manipulate the type of magnetic spiral, namely the angle between $\bbm{m}$ and $\bbm{k}$, by rotating the in-plane magnetization with respect to the lattice. 

The minimization of $\la f \ra $ as the function of $\bbm{k}$ gives,
\be
\bar{f} =\frac{A}{\mathcal{D}^{2}}\,\min_\bbm{k} \la f \ra =-\frac{9}{64} u_1 +\frac{AK}{2\mathcal{D}^2} u_2 -\frac{A}{\mathcal{D}^2}\cos\alpha\; \bbm{H}\cdot\bbm{m},
\e
where we have defined the functions
\beml
\begin{align}
u_1(\alpha, \theta) &= (\sin\alpha+5\sin 3\alpha)^2\sin^4\!\theta\,\cos^2\!\theta, \\  
u_2(\alpha, \theta) &= 2\cos^2\!\theta\,\cos^2\!\alpha+\sin^2\!\theta\,\sin^2\!\alpha.
\end{align}
\eml
The minimum of $\bar{f}$ is achieved either at $\alpha=0$ (collinear ferromagnetic state) or at a finite $\alpha$ (non-collinear spiral state). 

The function $\bar{f}$ should be additionally minimized with respect to the angles, $\alpha$, $\varphi$ and $\theta$. Since the azimuthal angle $\varphi$ enters $\bar{f}$ only via the Zeeman term $\bbm{H}\cdot\bbm{m}$, the in-plane magnetization $\bbm{m}_{\|}$ is always aligned with the in-plane magnetic field $\bbm{H}_{\|}$, which sets the angle $\phi$ at the minimum. The minimization with respect to $\alpha$ and $\theta$ remains non-trivial but is readily performed numerically. 

Thus, according to Eq.~\eqref{eq:dir}, one can manipulate the type of the spiral from $\bbm{k}\perp \bbm{m}_\parallel$ to $\bbm{k}\parallel \bbm{m}_\parallel$ by changing the direction of the in-plane magnetic field by the angle $\pi/6$. Such behavior is a signature of the quartic asymmetric exchange of the type of Eq.~(\ref{chiral}) in the non-collinear spiral phase.

\begin{figure}[!h]
\includegraphics[width=0.9\columnwidth]{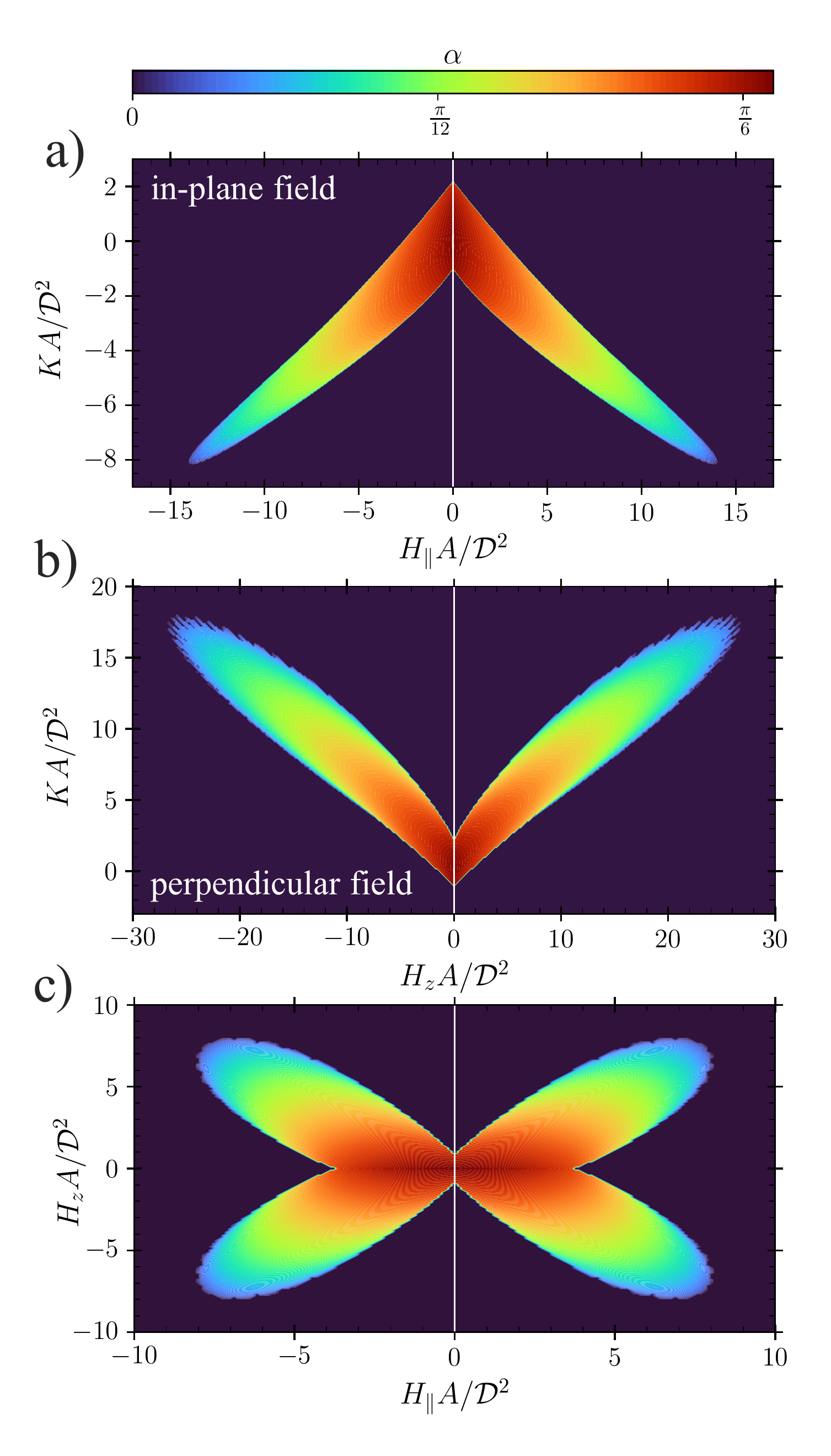}
\caption{Phase diagram of the collinear (black) and conical phases (colored). Panels (a) and (b): dependence on anisotropy $K$ and external magnetic field in lateral and perpendicular directions. correspondingly. 
Panel (c): the phase diagram in $(H_{\|},H_z)$ plane for vanishing anisotropy $K=0$.}	
\label{fig:phaseD}	
\end{figure}

The result of minimization of $\bar{f}$ for a general direction of the field $\bbm{H}=(\bbm{H}_{\|},H_z)$ leads to the phase diagram depicted schematically in Fig.~\ref{fig:phaseD}.  One can see that the in-plane magnetic field stabilizes the conical state for the case of easy axis anisotropy ($K<0$) as shown in Fig.~\ref{fig:phaseD}(a). Large perpendicular field may additionally stabilize the conical state for the easy plane anisotropy ($K>0$) as shown in Fig.~\ref{fig:phaseD}(a). 

%%%%%%%%%%%%%%%%%%%%%%%%%%%%
%%%% fig:phaseD
%%%%%%%%%%%%%%%%%%%%%%%%%%%%
%%%%%%%%%%%%%%%%%%%%%%%%%%%%

In Fig.~\ref{fig:phaseD}(c) we plot the phase diagram in $(H_{\|},H_z)$ plane for the case of vanishing anisotropy.  In this case, the opening angle of the cone reaches its maximal value $\alpha \approx \pi/6$ for the field tilted at 45 degrees to the plane, $H_{\|}=H_z$.  At the same time, there always exists a critical strength of the external field, beyond which the magnetic cone cannot be stabilized at any anisotropy value.

As one can see from the phase diagram of Fig.~\ref{fig:phaseD} the quartic asymmetric exchange must be sufficiently strong, $\mathcal{D}^2 \gtrsim A K$, in order to achieve a spiral ground state. Still, the presence of quartic asymmetric interaction can be detected in the collinear phase by studying the magnon spectra for various directions of the in-plane field. 

Let us now investigate magnon spectra for the collinear (ferromagnetic) phase. The dynamics of the magnetization vector $\bbm{n}(\textrm{r},t)$ is governed by the Landau-Lifshitz equation
\be
\frac{d\bbm{n}}{dt}=\bbm{n}\times \bbm{H}_\textrm{eff},
\label{LL}
\e
where $\bbm{H}_\textrm{eff}=-\delta F/\delta \bbm{n}$. 

Magnon spectra are obtained by linearization of Eq.~(\ref{LL}) around the ground state $\bbm{n}(\textrm{r},t)=\bbm{m}$, where $\bbm{m}$ is found from the minimization of the free energy density 
\be
f= K\,m_z^2-\bbm{H}\cdot \bbm{m}+\lambda\,H\,(m^2-1).
\e
Here $m=|\bbm{m}|$ and $\lambda$ stands for the Lagrange multiplier.

For $\bbm{H}=H\,(\sin\theta_\textrm{H}\cos\varphi_\textrm{H},\sin\theta_\textrm{H}\sin\varphi_\textrm{H},\cos\theta_\textrm{H})$ we find  
\be
\bbm{m}=\left( \frac{\sin\theta_\textrm{H}\cos\varphi_\textrm{H}}{2\lambda}, \frac{\sin\theta_\textrm{H}\sin\varphi_\textrm{H}}{2\lambda},\frac{\cos\theta_\textrm{H}}{2(\lambda+K/H)}\right),
\e
where $\lambda$ is obtained from the condition $|\bbm{m}|=1$. 

Thus, the azimuthal angles of in-plane magnetization and external field are always the same, $\varphi=\varphi_\textrm{H}$, while the polar angles differ $\sin \theta_\textrm{H}=2\lambda \sin\theta$ for the case of finite anisotropy. In the absence of anisotropy $\lambda=1/2$ and $\theta=\theta_\textrm{H}$ as well. 

Using $\bbm{n}(\bbm{r},t)=\bbm{m}+(\bbm{m}\times \bb{\delta})\,e^{i\bbm{q}\cdot\bbm{r}-i\omega t}$ and linearizing Eq.~(\ref{LL}) with respect to $\bb{\delta}$, we obtain the magnon spectra
\be
\omega^{\pm}_{\bbm{q}}=\pm \mathcal{S}(q^2,\lambda)+48\, \mathcal{D}q\,\sin^2\theta\,\cos\theta\,\sin(3\varphi+\chi),
\label{branches}
\e
where $\chi$ is the angle between the direction of magnon wave vector $\bb{q}$ and in-plane component of magnetization $\bb{m}_{\|}$. 

The second term in Eq.~(\ref{branches}) is absent if magnetization $\bbm{m}$ is directed perpendicular to the plane, $\theta=0$.  

The symmetric part of the magnon dispersion is given by
\be
\mathcal{S}(q^2,\lambda)=\sqrt{(\epsilon_q+K)(\epsilon_q+K\cos^2\theta)},
\e
where $\epsilon_q=A\,q^2+2\lambda\, H$. Note that two magnon branches are related by $\omega^{+}_{\bbm{q}}=-\omega^{-}_{-\bbm{q}}$.

\begin{figure}[!h]
\includegraphics[width=0.95\columnwidth]{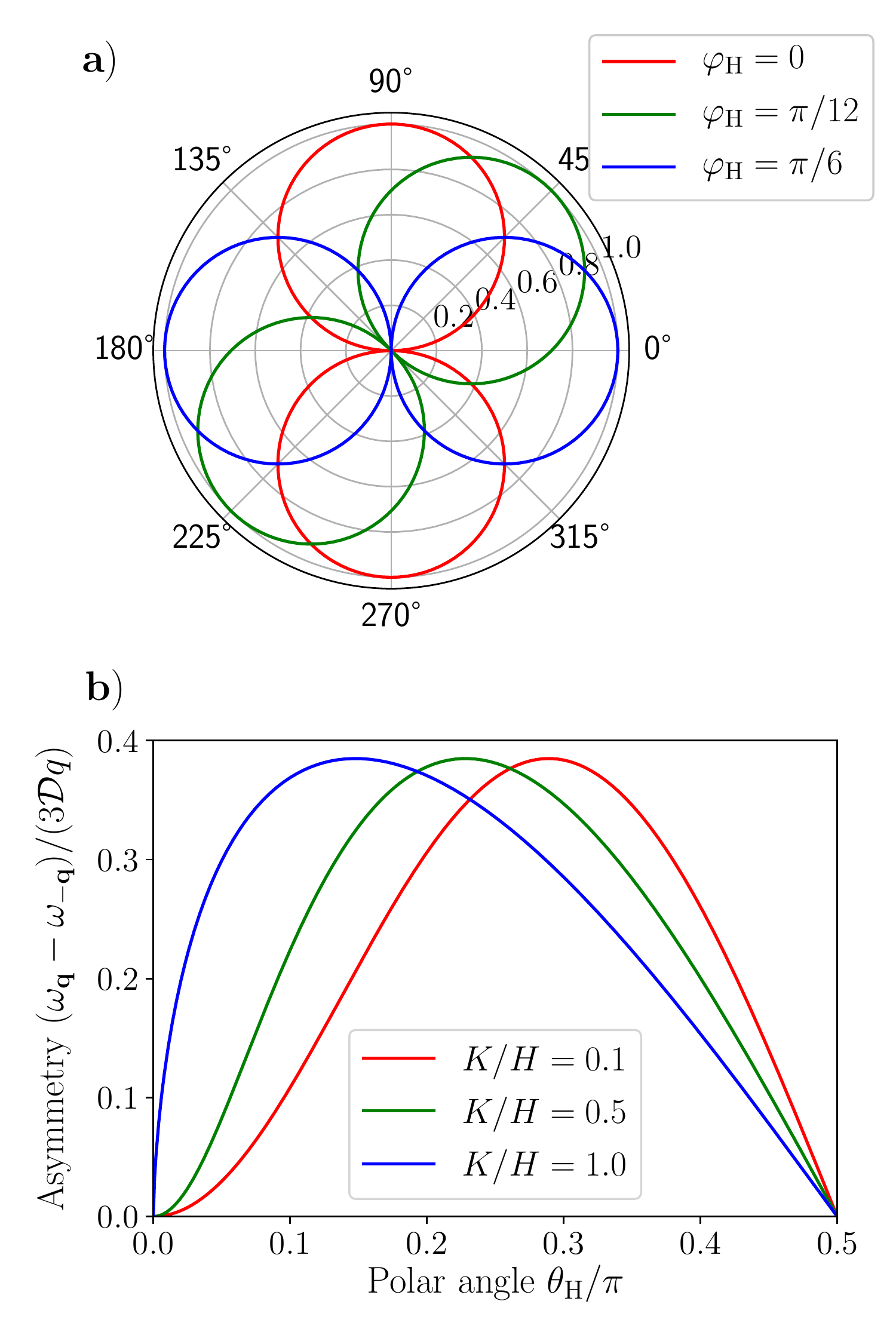}
\caption{Magnon branch asymmetry as a function of azimuthal $\varphi_\textrm{H}$ and polar $\theta_\textrm{H}$ angles of external magnetic field. Panel (a): the dependence on the relative angle $\chi$ between magnetic field and magnon propagation direction for three different values of $\varphi_\textrm{H}$ and $K/H=1$. Panel (b): the dependence on the polar angle $\theta_\textrm{H}$ for different vales of the relative anisotropy strength $K/H$ and $3\varphi+\chi=\pi/2$.}		
\label{fig:magnon} 
\end{figure}

The quantity $\delta\omega_\bbm{q}=\omega^{+}_\bbm{q}-\omega^{+}_{-\bbm{q}}$ characterizes the asymmetry between opposite magnon wave-vectors. The value of $\delta\omega_\bbm{q}$ is maximized for $\chi=\pm \pi/2-3\varphi$. Such an asymmetry is linear with respect to $\mathcal{D}q$ and has a non-trivial dependence on azimuthal and polar angles of the external magnetic field, $\varphi_\textrm{H}$ and $\theta_H$, as illustrated in Fig.~\ref{fig:magnon}. In the presence of anisotropy, the asymmetry can only be observed in oblique magnetic fields. 

The peculiar dependence of $\delta\omega_\bbm{q}$ on the direction of magnon propagation and on the direction of the in-plane magnetic field is specific for the asymmetric quartic exchange of Eq.~(\ref{chiral}). This dependence can be directly probed via recently demonstrated inelastic polarized neutron scattering~\cite{nambu2020observation}. The corresponding measurements can provide the directional profile of $\delta\omega_\bbm{q}$ shown in Figs.~\ref{fig:magnon}(a,b). Observation of the particular dependence of $\delta\omega_\bbm{q}$ on the direction of $\bbm{q}$ as shown in Fig.~\ref{fig:magnon}(a) would allow unambiguous attribution of the magnon asymmetry to the quartic asymmetric exchange of Eq.~(\ref{chiral}).  

%%%%%%%%%%%%%%%%%%%%%%%%%%%%
%%%% fig:magnon
%%%%%%%%%%%%%%%%%%%%%%%%%%%%
%%%%%%%%%%%%%%%%%%%%%%%%%%%%

To conclude, we investigated signatures of quartic asymmetric exchange in magnetic spiral phase and in magnon spectra of two-dimensional ferromagnets with $D_\textrm{3h}$ point group symmetry. Choosing in-plane component of magnetic field along different crystallographic directions changes the type of magnetic spiral in the non-collinear phase. Namely, the magnetic cone axis and the spiral propagation direction can be tuned from parallel to perpendicular orientation. In the collinear phase the rotation of in-plain magnetic field with respect to the lattice results in a strong modulation of magnon branch asymmetry. The latter effect can be probed experimentally e.\,g. via the inelastic polarized neutron scattering. Such measurements can unambiguously identify the quartic asymmetric exchange and characterize its strength. 

We are grateful to Marcos H. D. Guimar\~{a}es and Alexander Rudenko for numerous discussions. This research has been supported by the EU JTC-FLAGERA Project GRANSPORT.

%========================================================
%========================================================

\bibliographystyle{apsrev4-1}
\bibliography{Biblio}

\end{document}